\documentclass[aps,prl,reprint,showpacs,preprintnumbers,amsmath,amssymb,citeautoscript, superscriptaddress]{revtex4-1} 

\usepackage{graphicx}
\graphicspath{{figs/}{./}}
\usepackage{floatrow}
\usepackage[outercaption]{sidecap} 
\usepackage{color}
\newcommand\FigureFile[1] {#1.eps}
\DeclareGraphicsExtensions{pdf}

\usepackage{dcolumn}
\usepackage{bm}
\usepackage{color}
\usepackage{multirow}

\newcommand\eq[1]                              
{
\begin{align*}
#1
\end{align*}
}

\newcommand\eql[2] 
{
\begin{equation}\label{#1}
\begin{split}
#2
\end{split}
\end{equation}
}

\newcommand\eqll[3] 
{
\begin{equation*}\label{#1}
#3
\end{equation*}
}

\newcommand\eqsl[1]                            
{
\begin{align}
#1
\end{align}
}

\newcommand\eqssl[2]                      
{
\begin{subequations}\label{#1}
\begin{align}
#2
\end{align}
\end{subequations}
}

\definecolor{xmgrace-green4}{rgb}{0.0,0.55,0.0}
\definecolor{Green}{rgb}{0.2,0.96,0.2}
\definecolor{Remarks}{rgb}{1,0.3,0.3}
\definecolor{Extra}{rgb}{0.2,0.2,1}
\definecolor{Blue}{rgb}{0.2,0.3,1}
\definecolor{Black}{rgb}{0,0,0}

\newcommand\COMMENTED[1] {}

\begin{document}

\title{A Diamagnetic Trap with 1D Camelback Potential}

\author{Oki Gunawan}
\email{Correspondence: ogunawa@us.ibm.com}
\affiliation{IBM T. J. Watson Research Center,  Yorktown Heights, NY 10598}
\author{Yudistira Virgus}
\affiliation{IBM T. J. Watson Research Center,  Yorktown Heights, NY 10598}
\affiliation{Department of Physics, College of William and Mary, Williamsburg, VA 23187}
\date{\today}

\begin{abstract}
The ability to trap matter is of great importance in experimental physics since it allows isolation and measurement of intrinsic properties of the trapped matter. 
We present a study of a three dimensional (3D) trap for a diamagnetic rod in a pair of diametric cylindrical magnets. 
This system yields a fascinating 1D camelback potential along the longitudinal axis which is one of the elementary model potentials of interest in physics. 
This potential can be tailored by controlling the magnet length/radius aspect ratio. 
We develop theoretical models and verify them with experiments using graphite rods. 
We show that, in general, a camelback field or potential profile exists in between a pair of parallel linear dipole distribution. 
By exploiting this potential, we demonstrate a unique and simple technique to determine the magnetic susceptibility of the rod. 
This system could be further utilized as a platform for custom-designed 1D potential, a highly sensitive force-distance transducer or a trap for semiconductor nanowires.
\end{abstract}


\maketitle


Various matter and particle traps using optical or electromagnetic systems have been developed and instrumental in investigation of many physical phenomena \cite{Paul1990, Ashkin2000, Phillips1998}.
Most macroscale matter trap systems work for spherical or arbitrary shape objects \cite{Ashkin2000} 
but almost none has been specifically developed for cylindrical objects.
This work is initially motivated by the challenge to solve the problem of future electronic integrated circuit fabrication at the end of transistor scaling limit, specifically for semiconductor nanowire (or carbon nanotube) based integrated circuit \cite{ITRS2007, Appenzeller2008, Li200618}.
Such nanowire electronic circuit can be fabricated by top-down approach using conventional e-beam lithography \cite{Gunawan2008etal, Bangsaruntip2009etal}, however this method is expensive and has low throughput. 
An alternative technique is ``bottom-up" approach where the nanowires are grown, such as using vapor-liquid-solid technique \cite{Wagner1964} and then harvested in massive quantities \cite{Lu2008}.
Unfortunately there remains a key problem of how to assemble these nanowires precisely to targeted locations for integrated circuit fabrication. 
One possible route is to seek a scalable system that could trap cylindrical objects such as these nanowires. 
Many semiconductor materials including carbon nanotubes are diamagnetic \cite{Chadi1975, Lu1995}.
Such material will be attracted to a region with minimum magnetic field as has been demonstrated in  various magnetic levitation systems \cite{Geim1999, Ikezoe1998, Simon2000, Lyuksyutov2004, Kustler2007}.
Thus in principle, it should be possible to design certain magnetic configuration that can trap cylindrical diamagnetic objects.  

In this report, we study a 3D confinement produced by a pair of cylindrical \textit{diametric} magnets i.e. magnet with magnetization along the diameter. 
We discovered that a 1D camelback potential naturally arises along the longitudinal direction of the magnet. 
\emph{This potential is one of the elementary model potentials  of special interest in physics as it represents a simple confinement potential with two barriers}.
It is also reminiscent of a double rectangular barrier potential system that can be found in a resonant tunneling diode made of semiconductor double heterostructure \cite{Chang1974}. 
We investigate, both theoretically and experimentally, a macroscopic scale prototype utilizing cylindrical diametric magnets and graphite rods made of ordinary mechanical pencil leads [see Supplementary Information (SI) I]  as shown in Fig. 1.

\begin{figure}[bp]
\includegraphics[scale=0.59]{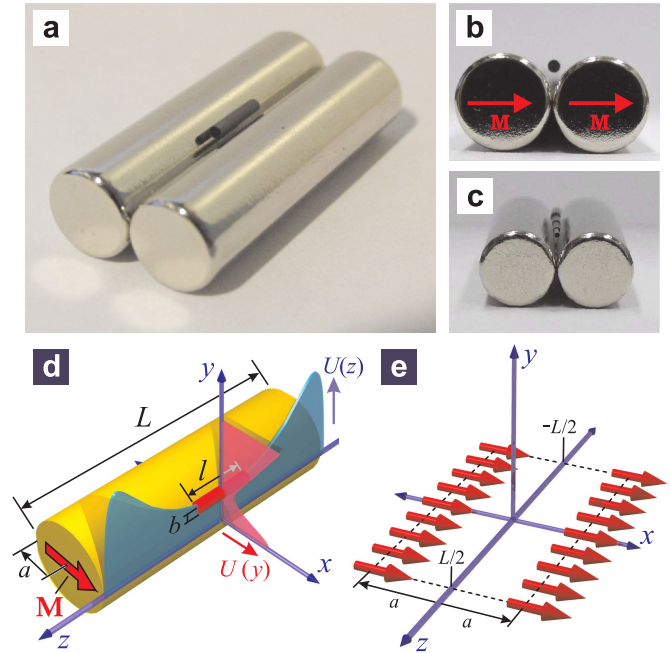}
\caption{\label{fig:Fig01} The diamagnetic trap with 1D camelback potential. \textnormal{
\textbf{(a)} The setup. 
\textbf{(b)} Cross section showing the magnet's magnetization $\mathbf{M}$. 
\textbf{(c)} Levitation of identical graphite rods of various diameters (0.3, 0.5, 0.7 and 0.9 mm).  
\textbf{(d)} Cross section showing the camelback $U(z)$ and vertical $U(y)$ confining potentials. 
\textbf{(e)} ``Parallel Dipole Line" model for magnetic field calculation at the center of the trap.}
}
\end{figure}

We will describe the magnetic field distribution from two cylindrical diametric magnets and the resulting confinement potentials.
First we consider a cylindrical diametric magnet centered at the origin with length $L$, radius $a$ and a uniform magnetization $M$ along $x$ axis: 
$ \mathbf{M} = M \mathbf{\hat{x}}$ as shown in Fig. 1d and S2a. 
The exact expression for the magnetic field (written in Cartesian vector form) can be derived using a magnetic scalar potential model (SI II.A.1): 
\begin{widetext}
\eql{eq:ScalarPot}
{   
       \mathbf{B}_{M}(x,y,z) = \frac{\mu_{0}Ma}{4\pi}\int_0^{2\pi}\sum_{n=1,2}\frac{(-1)^n}{u_{n}^2 + s^2 + u_{n}\sqrt{u_{n}^2 + s^2}}[x-a\cos\phi,y-a\cos\phi,u_{n} + \sqrt{u_{n}^2 + s^2}]\cos\phi \ \mathrm{d}\phi 
       \
}
\end{widetext}
where $\mu_{0}$ is the magnetic permeability in vacuum, $s^2 = (x-a\cos\phi)^2 + (y-a\cos\phi)^2$, $u_{1,2} = z \pm L/2$.
This expression has been verified experimentally (SI II.A.1) and at far distance $B_{M}$ approaches a pure dipole limit: $\mathbf{B}_{M}(x,0,0) \simeq \mu_{0}M a^2 L /2x^3 \mathbf{\hat{x}}$, which
is used to determine $M$.
We have also derived an alternative expression using magnetic vector potential (or bound surface current) model that gives identical result (SI II.A.2) but with separate contributions from the sheath and the end faces of the magnet. 

\begin{figure*}[tbp]
\includegraphics[scale=0.96]{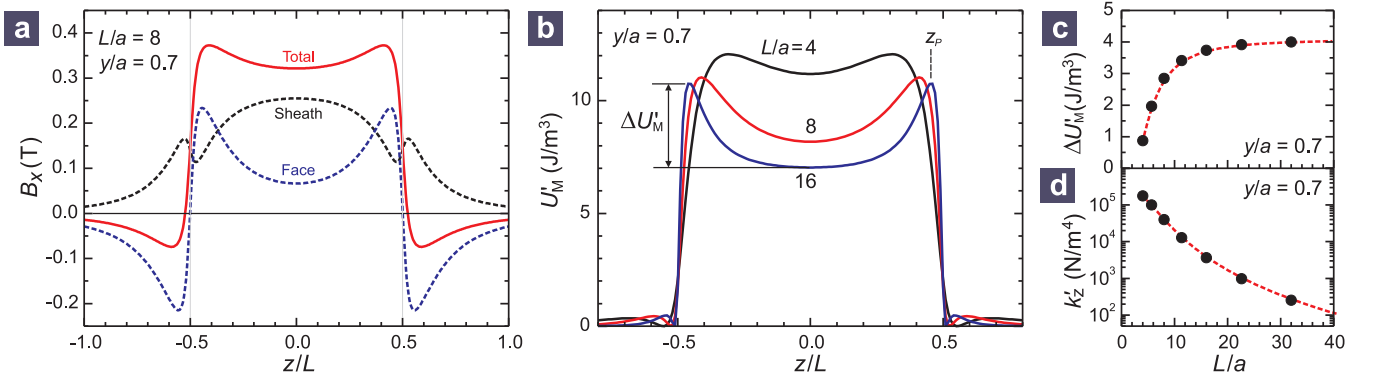}
\caption{\label{fig:Fig02} The camelback potential and its dependence on magnet aspect ratio $\textit{L/a}$ \textnormal{(at the center plane $x=0$). Magnet: $M = 10^{6}$ A/m, $a = 3.2$ mm. Rod: $\chi = -10^{-4}$.) 
\textbf{(a)} Magnetic field profile showing the ``sheath" ($B_{S}$) and ``face" ($B_{F}$) contributions and the total magnetic field ($B_{T}$). 
\textbf{(b)} The camelback potential profile for various $L/a$. $\Delta U_{M}^{'}$ and $z_{P}$ are the barrier height and peak position of the camelback hump. 
\textbf{(c)} Barrier height vs. $L/a$. 
\textbf{(d)} The potential ``spring constant" $k_{z}^{'}$ (per unit rod volume) is widely tunable by $L/a$.}
}
\end{figure*}

The trap has a pair of identical diametric magnets centered at  ($\pm{a}, 0, 0$) that naturally join and align their magnetizations in the same direction (Fig. 1b,d). 
The system will trap a diamagnetic rod with radius $b$, length $l$, mass density $\rho$ and magnetic susceptibility $\chi$  at the center plane ($x=0$) (Fig. 1d). 
The total magnetic field at the center plane is given as: $\mathbf{B}_{T}(y,z) = \mathbf{B}_{M}(a,y,	z) + \mathbf{B}_{M}(-a,y,z)$. 
Unfortunately $B_{M}$ (Eq.~\ref{eq:ScalarPot}) contains an integral with no analytic solution.

To facilitate simpler analysis we developed a ``\emph{parallel dipole line}'' (PDL) model, where the magnets are approximated by a distribution of magnetic dipoles in parallel lines at $x = \pm{a}$ along $-L/2 < z < L/2$ (Fig. 1e).
This model produces a closed form expression and a good approximation of the magnetic field at the center plane $x = 0$ (SI II.A.3): 
\eql{eq:PDL1}
{
       \mathbf{B}_{T}(y,z) = \frac{\mu_{0}M}{2(1+\bar{y}^2)^2}\sum_{n=1,2}\frac{\bar{w}_{n}[(1-\bar{y}^2)(\bar{y}^2 + \bar{w}_{n}^2) + 2]}{[1 + \bar{y}^2 + \bar{w}_{n}^2]^{3/2}} \mathbf{\hat{x}}
       \
}
where $\bar{y} = y/a$,  $\bar{w}_{1,2} = (L/2 \pm z)/a$.
Note that the magnetic field has only $x$ component due to the symmetry of the system.


We now investigate the vertical confining potential that levitates the graphite rod at the center of the trap ($x=0$, $z=0$). 
To focus on the essential physics, we use long magnet approximation ($\bar{L} = L/a \to \infty$) to Eq.~\ref{eq:PDL1} which yields: 
\eql{eq:PDL2}
{
       \mathbf{B}_{T\infty}(y,0) = \mu_{0}M \frac{1-\bar{y}^2}{(1+\bar{y}^2)^2} \mathbf{\hat{x}} 
       \
}
A cylindrical rod immersed in this magnetic field will have an induced magnetic moment $M_{R}$ given as (SI II.B): $\mathbf{M}_{R} = 2\chi\mathbf{B}_{T}/\mu_{0}(\chi + 2)$.
We assume a small rod radius ($b << a$) so that the magnetic field can be considered uniform over the radial extent of the rod.
Since the rod is a diamagnet, the induced magnetization is opposite to the magnetic field and tends to move it towards a region with minimum magnetic field  which leads to the levitation or trapping effect.
 
The stability condition can be investigated by considering the magnetic potential energy of the rod: $U_{M} = -V_{R}\mathbf{M}_{R} \cdot \mathbf{B}_{T}$, where  $V_{R} = \pi b^2 l$ is the rod's volume.
The total potential energy, including gravity, is given as: $U_{T}(y,z) = V_{R} (\rho gy -\mathbf{M}_{R} \cdot \mathbf{B}_{T})$, where $g$ is the gravitational acceleration. This potential provides a strong confinement in the vertical direction as illustrated in Fig. 1d (SI II.C).  The rod levitates at potential's minimum at position $\bar{y}_{0} = y_{0}/a$ which satisfies:
\eql{eq:PotEnergyMin}
{
      \rho g a + \mu_{0}M^2\chi/(\chi + 2) \times f_{Y}(\bar{y}_{0},\bar{L}) = 0 
              \
}
where  $f_{Y}(\bar{y}_{},\bar{L}) = -2a/\mu_{0}^2 M^2 \times \partial B_{T}^2(y,z)/\partial y$ is a dimensionless geometrical prefactor function proportional to the diamagnetic repulsion force in $y$-direction. 
Using the PDL model (Eq.~\ref{eq:PDL1}), we can calculate  $f_{Y}(\bar{y}_{},\bar{L})$ for any $\bar{L}$ (SI II.C). For the long magnet limit we obtain:  $f_{Y\infty}(\bar{y}_{})  = 8\bar{y}(3-\bar{y}^2)(1-\bar{y}^2)/(1+\bar{y}^2)^5$.

If $\chi$ is known, we can find the equilibrium height $y_{0}$ by solving Eq.~\ref{eq:PotEnergyMin}. 
Since both the diamagnetic repulsion and the gravity forces are proportional to the rod's volume, $y_{0}$ is independent of the rod's radius and length. 
Fig. 1c demonstrates this effect nicely where identical graphite rods of different diameters are aligned at the same height (see also Table S2). 
Analysis on the stability at the equilibrium point (SI II.C) implies the levitation only occurs at $y_{0}$ that satisfies: $0.287 < \bar{y}_{0} < 1$  with a minimum $|\chi|$ given as:
\eql{eq:ChiMin}
{
      |\chi|_{\textrm{min}} = 2/(1 + 4.136\mu_{0}M^2/\rho ga)
              \
}
Thus levitation can be more easily achieved with a rod that has stronger diamagnetic susceptibility and less density; and magnets with stronger magnetization but smaller radius.  
We also find that $|\chi|_{\textrm{min}}$ does not change significantly with varying $L/a$ (SI II.C).


We now investigate the confining potential along the longitudinal axis $z$. 
Using the ``Exact" scalar potential model (Eq.~\ref{eq:ScalarPot}) we can calculate the magnetic field profile. 
Furthermore, using the bound surface current (or magnetic vector potential) model described in SI II.A.2 we can calculate the individual contributions from the magnet's ``sheath" and end ``faces"  as shown in Fig. 2a. 
We observe that the ``humps" mainly arise from the ``face" contribution, or in other words, due to the finite length effect of the magnet. 
This can be intuitively understood from the bound surface current model (SI II.A.2): At the center plane only the $x$-component magnetic field  ($B_{x}$) exists and due to Biot-Savart law, any surface current in $y$ or $z$ directions will contribute to $B_{x}$. 
At the end faces, the bound current flows along the $y$ direction, thus near the edge of the magnet their contributions are stronger and gives rise to the camelback ``hump". 
We can express this camelback potential for a cylindrical diamagnetic rod levitated at height $y_{0}$ as: 
\eql{eq:UM}
{
      U_{M}^{'}(y_{0},z) = - 2\chi/\mu_{0}(\chi+2) \times B_{T}^{2}(y_{0},z)
              \
}
where  $U_{M}^{'} = U_{M}^{}/V_{R}$ is the energy potential per unit rod volume.
Using Eq.~\ref{eq:ScalarPot}, we can calculate this potential and the barrier height $\Delta U_{M}^{'}$ as shown in Fig. 2b. The camelback peak position can be estimated using PDL model as:  $z_{P} \simeq \pm(L/2 - \sqrt{2a^2 - y_{0}^2})$ (SI II.D).
Interestingly, we can tailor the shape of this camelback potential and the barrier height by tuning the magnet aspect ratio $L/a$ as shown in Fig. 2b,c. 

We investigated the longitudinal stability condition experimentally by cutting the graphite rods to various lengths. 
We find that for stable levitation, the length has to satisfy: $l_{\mathrm{min}} < l < l_{\mathrm{max}}$.
The maximum length is limited by the position of the camelback humps i.e. $l_{\mathrm{max}} \sim 2 z_{_P}$. 
Since the potential energy outside the humps drop very rapidly, the rod has to fit within the two humps to be trapped. 
The minimum length is caused by the fact that the magnetic field is mostly in $x$ direction.
Like a ferromagnetic rod, a diamagnetic rod also tends to align its longitudinal axis in the direction of the magnetic field  \cite{Puri1965}.
This effect is insignificant for a long rod $l > l_{\mathrm{min}}$
but when $l < l_{\mathrm{min}}$ the rod will align to $x$ direction, touches the surface of the magnets and no longer levitates. 
In our standard setup (SI I) we find that: $l_{\mathrm{min}} \sim a$ (see also Fig. S10).  A more quantitative analysis of $l_{\mathrm{min}}$ is a subject of further study.

In general, the confinement in the camelback potential along $z$-axis is significantly weaker compared to other directions ($x$ and $y$) (SI II.E). 
As a result, upon slight disturbance, the rod will oscillate as shown in Fig. 3a (see also Movie S1) with relatively long period $T_{z} \sim 1.4$s. 
To analyze the oscillation, besides assuming a small rod radius we also use short rod approximation ($l <<L$) and a small oscillation amplitude ($<<L$) so that the camelback potential at the center can be well approximated by a parabolic potential: $\Delta U_{z} (z) = \tfrac{1}{2}k_{z}z^2$ where $k_{z} = \partial U_{T}^2(0,y_{0},0)/\partial z^2$ is the harmonic potential ``spring constant" (SI II.F):
\eql{eq:SpringConstant1}
{
       k_{z} = -V_{R} \mu_{0}M^2 \chi/(\chi + 2) \times f_{Z2}(\bar{y}_{0},\bar{L})/L^2
        \
}
with $ f_{Z2}(\bar{y}_{},\bar{L}) =  2L^2/\mu_{0}^2M^2 \times \partial ^2 B_{T}^2(y,0)/\partial z^2$  is a dimensionless geometrical prefactor function for $k_{z} $. 
It can be calculated exactly using the exact model (Eq.~\ref{eq:ScalarPot}) or with the PDL model that yields (SI II.F): $f_{Z2}(\bar{y}_{},\bar{L}) = 192\bar{L}^4(\bar{L}^2 + 4\bar{y}^2 -16)[8 + (\bar{L}^2 + 4\bar{y}^2)(1 - \bar{y}^2)]/[(1 + \bar{y}^2)^2(4 + \bar{L}^2 + 4\bar{y}^2)^5]$.
This ``spring constant" $k_{z} $ can be widely tuned by the magnet aspect ratio $L/a$, for example, by a factor of $10^{-3}$ by changing $L/a$ from $4$ to $40$ (see Fig. 2d).

\begin{figure*}[tbp]
\floatbox[{\capbeside\thisfloatsetup{capbesideposition={right,top},capbesidewidth=4.4cm}}]{figure}[\FBwidth]
{\caption{Magnetic susceptibility determination analysis and experimental data for graphite rods of various  diameters. \textnormal{(See SI I for trap parameters). 
\textbf{(a)} An underdamped oscillation of a rod extracted from video footage (see lower inset and Movie S1).
 \textbf{Upper inset:} Damping time constant $\tau$ vs. rod diameter. 
 \textbf{(b)} Relationship between rod's $\chi$ and the equilibrium height $y_{0}/a$ using the ``Exact"  and the ``PDL"  models for setup $L/a = 4$ and $8$. 
 A data point of a HB/0.5 rod is shown. \textbf{Inset:} Diagram showing $y_{0}$. Shaded regions indicate unstable or no levitation for $L/a = 8$ setup. 
\textbf{(c)} Relationship between  period $T_{z}$ vs. the equilibrium height $y_{0}/a$ and experimental data for rod with various diameters. 
\textbf{(d)} Relationship between $\chi$ vs. $T_{z}$ and experimental data.}}\label{fig:Fig03}}
{\includegraphics[scale=0.91]{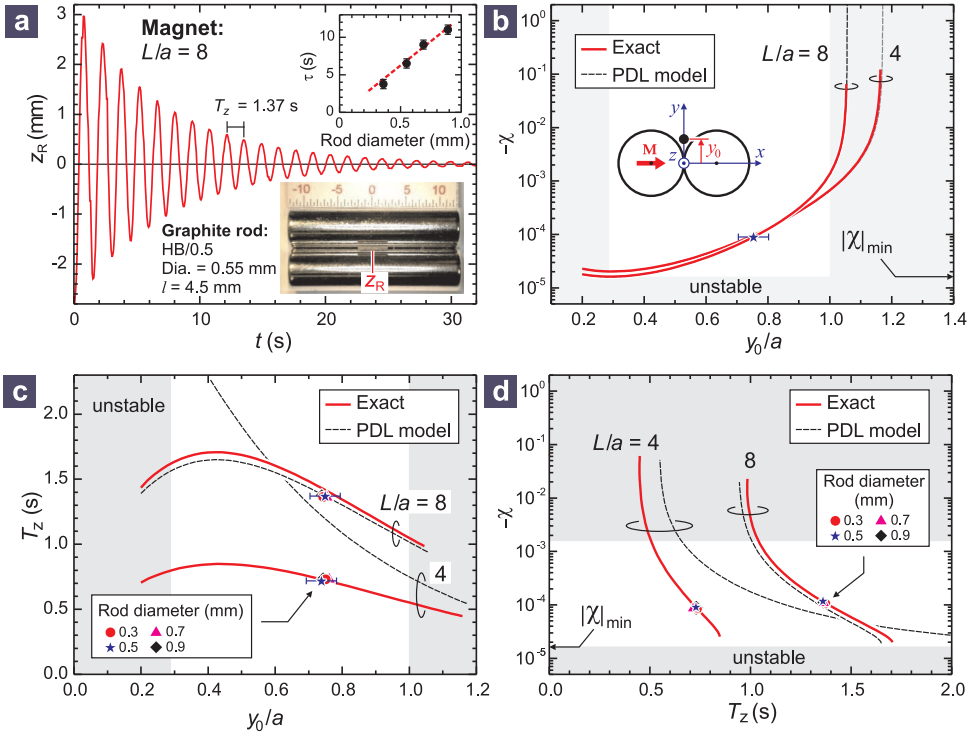}}
\end{figure*}

The oscillation period for the trapped rod can be expressed as: $T_{z}=2\pi \sqrt{m/k_{z}}$, where $m$ is the rod's mass. 
This leads to an interesting outcome where by measuring $T_{z}$, we could determine the rod's magnetic susceptibility given as (SI II.F):
\eql{eq:Chi2}
{
       \chi = - \frac{2}{1 + \mu_{0}M^2f_{Z2}(\bar{y}_{0},\bar{L})T_{z}^2/4\pi^2\rho L^2}
        \
}
Note that here we need to know $y_{0}$.
Surprisingly, $T_{z}$ is directly related to $y_{0}$ only by the geometrical factors of the magnet ($L$ and $a$) and independent of the magnetization $M$ and the property of the rod ($\rho$, $b$ and $l$).
This relationship is given below (SI II.F) and plotted in Fig. 3c:
\eql{eq:Periodz}
{
       T_{z} = f_{T}(y_{0},a,L) = 2\pi \sqrt{\frac{L^2 f_{Y}(\bar{y}_{0},\bar{L})}{gaf_{Z2}(\bar{y}_{0},\bar{L})}}
              \
}
Therefore to determine $\chi$ of the rod, we first measure $T_{z}$, solve for $y_{0}$ i.e. $y_{0} = f_{T}^{-1}(T_{z},a,L)$ and then use Eq.~\ref{eq:Chi2}.  
 We illustrate the $\chi$ measurement in two magnet trap setups with aspect ratio: $L/a = 4$ and $8$ but the same radius ($a = 3.2$ mm, see SI I) using a short graphite rod as presented in Fig. 3. 
 We provide two calculation models: ``Exact" i.e. using the exact magnetic field formula (Eq.~\ref{eq:ScalarPot}) and ÒPDL modelÓ (Eq.~\ref{eq:PDL1}).   
First we measure  $T_{z}$ and  $y_{0}$ then plot the data points in Fig. 3c. 
We also plot the expected $T_{z}$ vs. $y_{0}$ curves from Eq.~\ref{eq:Periodz}. 
We observe good agreement between the data and the ``Exact" model for both magnet setups, therefore given $T_{z}$ we could also determine $y_{0}$ without measuring it.

Next we determine $\chi$ from Eq.~\ref{eq:Chi2} using the ``Exact" model, as plotted in Fig. 3d. 
Measurements from both setups yield good agreement i.e. $\chi = -(11.0\pm1.6)\times10^{-5}$ and  $\chi = -(9.0\pm1.9)\times10^{-5}$ for setup $L/a = 8$ and $4$ respectively indicating the consistency of our model. 
Note that these results are within the reported $\chi$ value for graphite in literature:  $\chi_{\perp} = -1.4 \times10^{-5}$ and $\chi_{\parallel}  = -61 \times 10^{-5}$ \cite{Ganguli1941} for $\chi$ measured along perpendicular and parallel to the $c$-axis respectively. 
Our graphite pencil is amorphous thus its $\chi$ should be a mixture of both $\chi$ orientations. 
 
One could also determine $\chi$ from $y_{0}$ as illustrated in Fig. 3b, however, $T_{z}$  measurement is easier and more accurate (unlike for $y_{0}$, error bars for  $T_{z}$ are small and not visible in Fig. 3). 
Fig. 3b-d also show that the PDL model becomes closer to the ``Exact" model only for long magnet case ($L/a \gtrsim 5$).
This is reasonable as the PDL model provides better approximation for longer magnet (see Fig. S5).
We have also investigated diameter and length dependence effect of the rod.
Fig. 3c-d show that different diameters ($0.3 - 0.9$ mm) yield identical results which is expected from our small rod radius approximation (i.e. rod diameter has no effect). 
The effect of rodÕs length and short rod approximation is discussed in SI II.F.

Finally, we observe that the oscillation is underdamped following: $z_{R} \propto \exp(-t/\tau)\sin\omega t$
 , where $\tau$ is the damping constant, $\omega = 2\pi/T_{z}$ and $t$ is time. 
Note that a severe damping ($\tau \lesssim T_{z}$) could artificially increase the measured $T_{z}$ (SI II.G).  
This damping could be due to air friction (viscosity) or eddy current braking effect.
The latter could be a significant effect for a conductor 
\footnote{Graphite is a sufficiently good conductor with conductivity $\sim 1/300$ of copper.}
moving in a strong magnetic field. 
To determine the main cause of the damping we perform experiments in vacuum and with different rod diameters (SI II.G).
We observe that the damping gets weaker in vacuum and $\tau$ is proportional to the rodÕs diameter (see Fig. 3a inset) which are signatures of air friction effect. 
Thus both tests confirm that the damping effect is mainly due to air friction.  

 
 In summary, we show that a pair of diametric magnets provides a 3D trap for a diamagnetic rod and produces a fascinating 1D camelback potential along the longitudinal axis. 
 The potential humps arise mainly due to the end faces contribution of the magnetÕs surface current or due to finite size effect of the magnet.
 In general, \emph{we show that a diamagnetic camelback potential will arise at the center of a parallel linear dipole distribution} as described by our PDL model (Fig. 1e). 
 The shape of this potential can be tailored by adjusting the magnetsÕ aspect ratio $L/a$.
We have developed theoretical models that describe the magnetic field distribution, the potential trap profile, the stability condition and the oscillation dynamics along the longitudinal axis. 

A potential system which is tunable in space and time and particularly with reduced dimension is of special interest in physics and this trap system could serve such a purpose. 
By joining segments of magnet pairs with different magnetization one could realize almost any arbitrary 1D potential (see SI II.H). 
Similarly one could use electromagnet to achieve temporal control of the potential. 
Due to its simple configuration, this system is scalable to various length scales and may find different applications in different regimes. 

In small scale, the system could be utilized to trap semiconductor nanowires as originally intended in this study \footnote{O. Gunawan, Q. Cao, Magnetic trap for cylindrical diamagnetic materials. US Patent Application, IBM Docket YOR920120063US1 (2013).}.
Our model (Eq.~\ref{eq:ChiMin}) indicates that trapping is easier to achieve at smaller scale (smaller $a$) which is important as the diamagnetism in semiconductors are weaker than graphite \cite{Chadi1975}.
In macroscopic scale, we have demonstrated a simple magnetic susceptibility measurement of a trapped rod which is much simpler compared to other existing techniques (e.g. vibration sample magnetometer \cite{Foner1959}), provided the material-under-test can be prepared in cylindrical rod form. 
Furthermore, the damping time constant of the oscillation can be utilized to extract the viscosity of the ambient gas.  
This system can also be utilized as a highly sensitive force-distance transducer whose spring constant is widely tunable by the magnetÕs aspect ratio $L/a$ (Fig. 2d). 
A tiny force can be coupled to the trapped rod and the displacement provides a force read-out. 
Finally, as this system can be easily realized in macroscopic scale, it provides a fascinating pedagogical example to demonstrate the physics of diamagnetic levitation and ``particle in a 1D camelback potential" system.

We thank Chang Tsuei and Qing Cao (IBM) and Andika Putra (University of Maryland) for their constructive review and suggestions.
 Supplementary Information (SI) and Movie S1 are available online.
 \bibliography{MAGTRAP}

\end{document}